\documentclass[twocolumn,pre,nobibnotes]{revtex4}
\usepackage{graphicx,amsmath,bm,color,epstopdf}

\begin{document}

\title{Approximation methods for the stability of complete synchronization on duplex networks}
\author{Wenchen Han}
\affiliation{School of Science, Beijing University of Posts and
Telecommunications, Beijing, 100876, People's Republic of China}
\author{Junzhong Yang}\email{jzyang@bupt.edu.cn}
\affiliation{School of Science, Beijing University of Posts and Telecommunications,
Beijing, 100876, People's Republic of China}

%\date{\today}

\begin{abstract}
Recently, the synchronization on multi-layer networks has drawn a lot of attention.
In this work, we study the stability of the complete synchronization on duplex networks.
We investigate effects of coupling function on the complete synchronization on
duplex networks. We propose two approximation methods to deal with the stability
of the complete synchronization on duplex networks. In the first method, we introduce
a modified master stability function and, in the second method, we only take into
consideration the contributions of a few most unstable transverse modes to the
stability of the complete synchronization. We find that both methods work well
for predicting the stability of the complete synchronization for small networks.
For large networks, the second method still works pretty well.
\end{abstract}

\maketitle

\section{Introduction}

Network science has provided a fertile ground for understanding
complex systems. The traditional network approach treats complex systems
as monolayer networks by charting an elementary unit into a network node
and representing each unit-unit interaction on an equivalent footing as
a network link. \cite{boc06,new10,sza07,pas15,boc02,cas09,bull12}.
Recently, it has become clear that many complex systems in social,
biological, technological systems should not be treated as monolayer ones
but multi-layer ones \cite{zan15,boc14,dom13,kiv14,bul10,dom16,gra13,
gom12,agu14,sor12}.
Consider a group of members interacting with each other through different
channels such as Twitter, blog, Facebook, and Wechat. The social network
formed by these members is a typical example of multilayer network in
which each interaction channel corresponds to a different layer.

Synchronization, one of the most interesting collective behaviors, has been
investigated since the dawn of natural science \cite{boc02,boc16,alex08,pik03,lin16}.
There are different types of synchronization such as complete
synchronization (CS) \cite{boc02}, phase synchronization \cite{ros96},
lag synchronization \cite{ros97}, general synchronization \cite{rul95},
cluster synchronization \cite{bel01}, partial synchronization \cite{vre96},
and remote synchronization \cite{gam16}. Among all types of synchronization,
CS , where states of all oscillators are identical, is the simplest one \cite{sch16}.
CS on monolayer networks has been well studied by applying the master stability
function (MSF) method \cite{pec98,yang98}. The MSF method shows that the
stability of CS is affected by coupling functions (CFs) and network structures.

Recently, much attention has been paid to synchronization on multi-layer networks
\cite{boc14,sor12,agu14,bil14,kra14,sev15,li16,gen16}. Aguirre \textit{et al.} have
shown that connecting the high-degree (or low-degree) nodes in different layers
turns out to be the most (or the least) effective strategy to achieve synchronization
in multi-layer networks \cite{agu14}.
Using the MSF method, Sorrentino \textit{et al.} have studied CS on duplex networks,
one special type of two-layer networks with the two layers sharing the same
nodes \cite{boc14}, when the two layers are subject to constrains such as
commuting Laplacians, unweighted and fully connected layers, and nondiffusive
coupling \cite{sor12}.
Genio \textit{et al.} have provided a full mathematical framework to evaluate the
stability of CS on multi-layer networks by generalizing the MSF method \cite{gen16}.
However, $N-1$ transverse modes are coupled in their framework and the stability
of CS is hard to deal with for large $N$. Then, one question arises:
can we develop approximation methods to reduce the stability of CS on
multi-layer networks to a low dimensional problem?

In this work, we study the coupled identical chaotic oscillators on duplex networks.
We develop two approximation methods to deal with the stability of CS on duplex networks.
In the first method, we assume that all transverse modes to synchronous chaos
have the same contribution to the stability of CS. Then we obtain a modified MSF
similar to the MSF on monolayer networks. In the second approximation,
we only consider the contributions from a few most unstable transverse modes
of layers, which are responsible for the desynchronization on each isolated layer,
to the stability of CS.
The stability diagrams of CS on duplex networks produced by the two approximation
methods are compared with the direct numerical results on coupled chaotic
oscillators and we find that the second approximation method provides better
prediction on the stability of CS when $N$ is large.

\section{The Model}

We consider N oscillators sitting on a duplex network, whose time evolution
is governed by
\begin{equation} \label{eq1}
\dot{\bm{x}}_i=\bm{F}(\bm{x}_i)+\sum_{j=1}^{N}{[\varepsilon^{(1)} L^{(1)}_{i,j}
\bm{H}^{(1)}(\bm{x}_j)+\varepsilon^{(2)}L^{(2)}_{i,j}\bm{H}^{(2)}(\bm{x}_j)]},
\end{equation}
where $\bm{x}_i$ is an $m$-dimensional state variable of node $i$, $\bm{F}(\bm{x})$
is the dynamics of individual nodes, $\varepsilon^{(1)}$ (or $\varepsilon^{(2)}$) is
the coupling strength  and $\bm{H}^{(1)}(\bm{x})$ [or $\bm{H}^{(2)}(\bm{x})$]
is a linear CF, determining the output signal from a node on the layer $1$
(or the layer $2$). $L^{(1)}$ (or $L^{(2)}$) is the Laplacian matrix of
the layer $1$ (or the layer $2$), with elements $L^{(1)}_{i,i}=-k^{(1)}_i$
(or $L^{(2)}_{i,i}=-k^{(2)}_i$), the degree of node $i$ on the layer $1$ (or the
layer $2$), $L^{(1)}_{i,j}=1$ (or $L^{(2)}_{i,j}=1$) if node $i$ and node $j$
are connected with a link on the layer $1$ (or the layer $2$), and $L^{(1)}_{i,j}=0$
(or $L^{(2)}_{i,j}=0$) otherwise.

We briefly review the MSF method on the stability of CS on monolayer networks
\begin{equation} \label{eq2}
\dot{\bm{x}}_i=\bm{F}(\bm{x}_i)+\varepsilon\sum_{j=1}^{N}L_{i,j}\bm{H}(\bm{x}_j)
\end{equation}
The variational equations of Eq.~\eqref{eq2} with respect to CS,
($\bm{x}_1=\bm{x}_2=\cdots=\bm{x}_N=\bm{s}$), are diagonalized into $N$ decoupled
eigenmodes of the form
\begin{equation} \label{eq3}
\dot{\bm{\eta}}_i=[\mathcal{D}\bm{F}(\bm{s})+\varepsilon\lambda_i
\mathcal{D}\bm{H}(\bm{s})]\bm{\eta}_i
\end{equation}
where $\lambda_i$ ($i=1,2,...,N$) are eigenvalues of the Laplacian matrix $L$
and $L\phi_i=\lambda_i\phi_i$. For an undirected network, where $L$ is symmetric,
$\lambda_i$ are real and can be sorted in descending order, i.e.,
$0=\lambda_1>\lambda_2\geq...\geq\lambda_N$.
The eigenmode $\phi_1$ with $\lambda_1=0$ accounts for the synchronous mode and
other $\phi_i$, for $i=2,3,...,N$, are transverse modes.
$\mathcal{D}\bm{F}(\bm{s})$ and $\mathcal{D}\bm{H}(\bm{s})$ are the $m\times m$
Jacobian matrices of the corresponding vector functions evaluated at CS.
Letting $\sigma=\varepsilon\lambda$, the largest Lypunov exponent (LLE) $\Lambda(\sigma)$,
determined from Eq.~\eqref{eq3}, is the MSF. Generally, $\Lambda(\sigma)$ is negative
when $\sigma_1<\sigma<\sigma_2$.
Therefore, CS is stable when all transverse eigenmodes with $i>1$ have negative LLE,
that is, $\sigma_1<\varepsilon\lambda_i<\sigma_2$ for any $i>1$.
For a given node dynamics, CF can be categorized according to $\sigma_1$ and $\sigma_2$.
There are three types of CF. For the type-i CF, $\sigma_1=\infty$ and CS is always
unstable. For the type-ii CF, $\sigma_2=\infty$ and CS is stable provided that
$\varepsilon>\sigma/\lambda_2$. For the type-iii CF, both $\sigma_1$ and $\sigma_2$
are finite and CS is stable when $\sigma_1/\lambda_2<\varepsilon<\sigma_2/\lambda_N$.

\begin{figure}
\begin{center}
\includegraphics[width=3.4in]{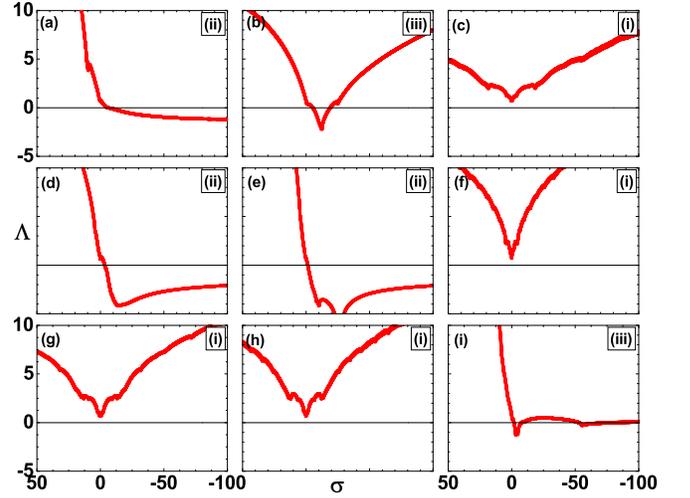}
\end{center}
\caption{\label{fig1}
The largest Lyapunov exponent $\Lambda$ (in red) against $\sigma$ for coupled
Lorenz oscillators on monolayer networks for different CFs.
(a) $\mathcal{D}\bm{H}_{1,1}$, (b) $\mathcal{D}\bm{H}_{1,2}$,
(c) $\mathcal{D}\bm{H}_{1,3}$, (d) $\mathcal{D}\bm{H}_{2,1}$,
(e) $\mathcal{D}\bm{H}_{2,2}$, (f) $\mathcal{D}\bm{H}_{2,3}$,
(g) $\mathcal{D}\bm{H}_{3,1}$, (h) $\mathcal{D}\bm{H}_{3,2}$, and
(i) $\mathcal{D}\bm{H}_{3,3}$. The index in the top-right corner in each plot
denotes the type of the corresponding CF.}
\end{figure}

In the following, we take chaotic Lorenz oscillator ($m=3$) as the node dynamics,
which is described as $\bm{F}(\bm{x})=[10(y-x),28x-y-xz,xy-z]^T$.
We concern with CFs whose Jacobian matrices have only one nonzero element and we
denote them with their nonzero element. Thereby, there are $9$ different CFs.
Fig.~\ref{fig1} shows the LLE, $\Lambda(\sigma)$, for $9$ different CFs.
The type of the CF in each plot is marked by the index in the top-right corner.

\section{Numerical Simulations}

In this section, we numerically investigate the dependence of CS on CFs in coupled
Lorenz oscillators on duplex networks with $N=6$. Both layers of duplex networks
are modeled by random networks. We have tried different realizations of duplex networks
and found qualitatively similar results. Without the loss of generality, we consider
a specific duplex network with the Laplacians,
$L^{(1)}=\left(\begin{array}{cccccc}-2&1&0&0&0&1\\1&-3&1&1&0&0\\
0&1&-4&1&1&1\\0&1&1&-3&0&1\\0&0&1&0&-2&1\\1&0&1&1&1&-4\\ \end{array}\right)$ and
$L^{(2)}=\left(\begin{array}{cccccc}-3&0&1&0&1&1\\0&-2&0&0&1&1\\
1&0&-3&0&1&1\\0&0&0&-1&1&0\\1&1&1&1&-5&1\\1&1&1&0&1&-4\\ \end{array}\right)$.
We consider the synchronization error, which is defined as
\begin{equation} \label{eq4}
\Delta=\frac{2}{N(N-1)}\sum\limits_{i=1,j>i}^N{\langle||\vec{x}_j-\vec{x}_i||_2}\rangle_t,
\end{equation}
where $\langle x\rangle_t$ means the time average of $x$ and
$||\vec{x}_j-\vec{x}_i||_2$  is the Euclidean norm
($||\vec{x}_j-\vec{x}_i||_2=\sqrt{\sum_{l=1}^{m}(x_{j,l}-x_{i,l})^2}$).
In each simulation, the synchronization error is averaged over $400$ time
units after a transient time around $2000$ time units.
When $\Delta<10^{-6}$, we say that the coupled Lorenz oscillators are
completely synchronized.

\begin{figure}
\begin{center}
\includegraphics[width=3.4in]{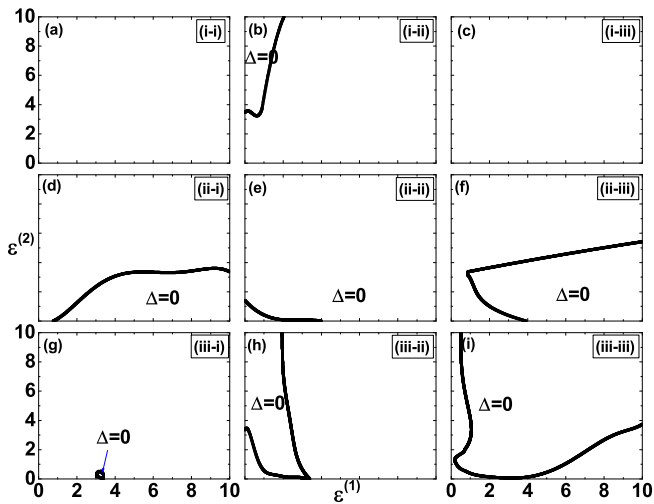}
\end{center}
\caption{\label{fig2} The stability diagram of CS on the $\varepsilon^{(1)}-
\varepsilon^{(2)}$ plane in coupled Lorenz system on duplex networks
for different combinations of CFs. (a) $\mathcal{D}\bm{H}^{(1)}_{2,3}=1$ and
$\mathcal{D}\bm{H}^{(2)}_{3,1}=1$, (b) $\mathcal{D}\bm{H}^{(1)}_{2,3}=1$ and
$\mathcal{D}\bm{H}^{(2)}_{2,1}=1$, (c) $\mathcal{D}\bm{H}^{(1)}_{2,3}=1$ and
$\mathcal{D}\bm{H}^{(2)}_{3,3}=1$, (d) $\mathcal{D}\bm{H}^{(1)}_{2,2}=1$ and
$\mathcal{D}\bm{H}^{(2)}_{1,3}=1$, (e) $\mathcal{D}\bm{H}^{(1)}_{1,1}=1$ and
$\mathcal{D}\bm{H}^{(2)}_{2,2}=1$, (f) $\mathcal{D}\bm{H}^{(1)}_{1,1}=1$ and
$\mathcal{D}\bm{H}^{(2)}_{1,2}=1$, (g) $\mathcal{D}\bm{H}^{(1)}_{1,2}=1$ and
$\mathcal{D}\bm{H}^{(2)}_{1,3}=1$, (h) $\mathcal{D}\bm{H}^{(1)}_{1,2}=1$ and
$\mathcal{D}\bm{H}^{(2)}_{2,1}=1$, (i) $\mathcal{D}\bm{H}^{(1)}_{1,2}=1$ and
$\mathcal{D}\bm{H}^{(2)}_{3,3}=1$.}
\end{figure}

As shown in Fig.~\ref{fig1}, all three types of CF are presented for Lorenz
oscillators on monolayer networks. Consider coupled Lorenz oscillators on the
duplex network in which two layers take CFs with different types. There are
$9$ typical combinations of CFs for duplex networks. Then we numerically
explore stable CS on the plane of $\varepsilon^{(1)}$ and $\varepsilon^{(2)}$
and investigate effects of combinations of different types of CFs on CS.
The results are presented in Fig.~\ref{fig2}, where the combinations of CFs
are labelled in each plot, for example (i-iii) indicating a type-i CF
on the layer $1$ and a type-ii CF on the layer $2$.

The results can be summarized as follows. Firstly, the presence of type-i CFs
always disfavors CS. When both layers take the type-i CF, it is impossible to
realize CS [see Fig.~\ref{fig2} (a)]. As shown in Fig.~\ref{fig2} (b) [or (d)] with
the combination of CFs taking (i-ii) [or (ii-i)], increasing the coupling strength
of the type-i CF shrinks the range of the coupling strength of the type-ii CF
supporting stable CS. The similar effects of the type-i CF on CS can be observed
for the combination of CFs taking (i-iii) in Fig.~\ref{fig2} (c) and (iii-i) in
Fig.~\ref{fig2} (g). Fig.~\ref{fig2} (c) suggests that, if a
type-iii CF does not support stable CS, stable CS cannot be built on
duplex networks with the combinations of CFs taking (i-iii) or (iii-i). Secondly,
the presence of type-ii CFs always enhances synchronization.
Increasing the coupling strength of the type-ii CF expands the
range of the coupling strength of the CF on the other layer allowing for the stable CS.
Especially for the combination (ii-ii) in Fig.~\ref{fig2} (e), only a small region
on the plane of $\varepsilon^{(1)}$ and $\varepsilon^{(2)}$ prohibits stable CS.
Thirdly, the presence of type-iii CFs makes the dependence of stable CS
on CFs more interesting. As shown in Fig.~\ref{fig2} (f) with the combination
of CFs as (ii-iii), even if a type-iii CF does not supprt CS on monolayer
networks, it might play a very positive role on the stable CS on duplex networks.
When (iii-iii) is taken by duplex networks [see Fig.~\ref{fig2} (i)],
the enhancement of synchronization by
the interplay between type-iii CFs on two layers is extraordinary strong.

\section{Analysis and Approximation methods}

\begin{figure}
\begin{center}
\includegraphics[width=3.4in]{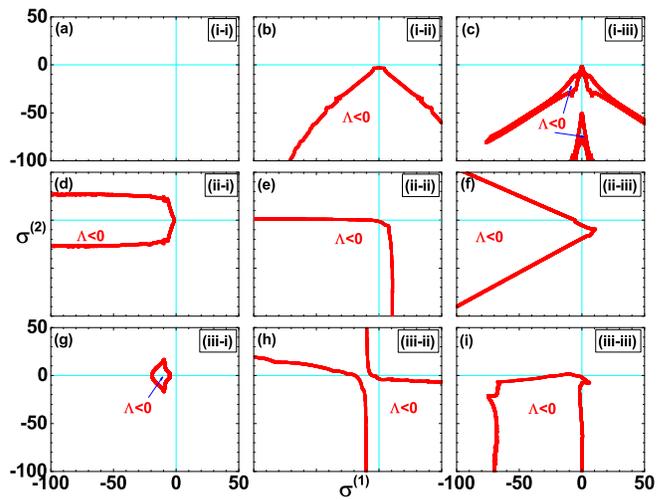}
\end{center}
\caption{\label{fig3}
The master stability function $\Lambda(\sigma^{(1)},\sigma^{(2)})$ calculated according
to Eq.~\eqref{eq6}. The cyan lines are for $\sigma^{(1)}=0$ and $\sigma^{(2)}=0$.
The CFs combinations are identical to those in Fig.~\ref{fig2}.}
\end{figure}

In this section, we theoretically investigate the stability of CS on duplex
networks. Similar to monolayer networks, the stability of CS on duplex networks
is determined by the variational equation of  Eq.~\eqref{eq1} with respect to CS
\begin{eqnarray} \label{eq5}
\frac{d}{dt}\bm{\eta}^{(\alpha)}_i&=&[\mathcal{D}\bm{F}(\bm{s})+
\varepsilon^{(\alpha)}\lambda^{(\alpha)}_i\mathcal{D}\bm{H}^{(\alpha)}
(\bm{s})]\bm{\eta}^{(\alpha)}_i \\ \nonumber &+&\varepsilon^{(\beta)}
\sum_{j=2}^{N}{\mathcal{L}^{(\beta,\alpha)}_{i,j}\mathcal{D}\bm{H}^{(\beta)}
(\bm{s})\bm{\eta}^{(\alpha)}_j},
\end{eqnarray}
with $\alpha,\beta=1,2$ and $\alpha\neq\beta$,
where $i=2,3,...,N$, $\lambda^{(\alpha)}_i$ are the eigenvaules of the Laplacian matrix
$L^{(\alpha)}$ ($L^{(\alpha)}\phi^{(\alpha)}_i=\lambda^{(\alpha)}_i\phi^{(\alpha)}_i$),
and $\mathcal{L}^{(\beta,\alpha)}=(\phi^{(\alpha)})^TL^{(\beta)}\phi^{(\alpha)}$.
Different from Eq.~\eqref{eq2}, where $N-1$ decoupled $m$-dimensional ordinary
differential equations (ODEs) are obtained, Eq.~\eqref{eq5} depicts $N-1$ coupled
$m$-dimensional equations. Consequently, the MSF method, which reduces a high-dimensional
[$N-1$ $m$-dimentional] stability problem to a low-dimensional ($m$-dimensional) one
on monolayer networks, cannot be applied to the stability of CS on duplex networks.
Only for some special cases where $\mathcal{L}^{(\beta,\alpha)}$ can be diagonalized,
Eq.~\eqref{eq5} is reduced to $N-1$ decoupled $m$-dimensional equations.
For example, duplex networks
in which the Laplacian matrices satisfy $L^{(1)}L^{(2)}=L^{(2)}L^{(1)}$ or duplex
networks with one layer whose Laplacian matrix has $N-1$ same eigenvalues except
for $\lambda_1=0$ \cite{sor12}. For arbitrary duplex networks, the analysis of
the stability of CS is not an easy job, especially for large $N$. In the following,
we propose two approximation methods to reduce Eq.~\eqref{eq5} to a low-dimensional
problem and make the stability analysis of CS easier on duplex networks.

\subsection{Approximation method I}

\begin{figure}
\begin{center}
\includegraphics[width=3.4in]{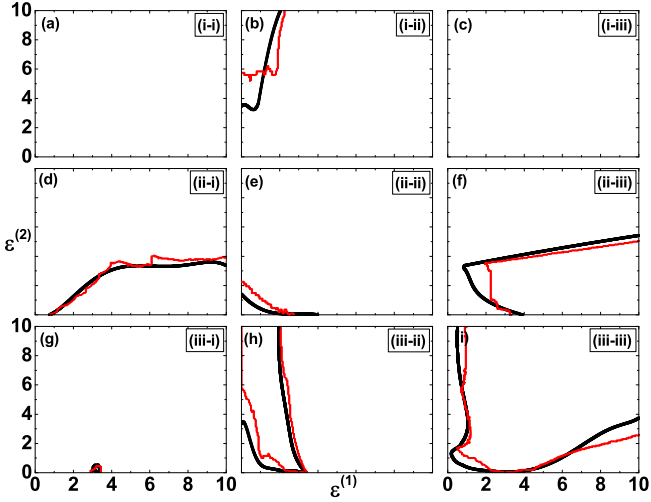}
\end{center}
\caption{\label{fig4}
The stable diagram of CS predicted by the approximation method I (in red)
and obtained by the synchronization error (in black).
The CFs are identical to those in Fig.~\ref{fig2}.}
\end{figure}

The MSF method is powerful in stability analysis of CS on monolayer networks.
Can we find a similar MSF for CS on duplex networks? For this aim, we have to
decouple the $N-1$ $m$-dimensional equations in Eq.~\eqref{eq5}.
A plausible way to do it is approximating the last term, $\sum_{j=2}^{N} {\mathcal{L}^
{(\beta,\alpha)}_{i,j}\mathcal{D}\bm{H}^{(\beta)}(\bm{s})\bm{\eta}^{(\alpha)}_j}$,
in Eq.~\eqref{eq5} with $\sum_{j=2}^{N}{\mathcal{L}^
{(\beta,\alpha)}_{i,j}\mathcal{D}\bm{H}^{(\beta)}(\bm{s})}\bm{\eta}^{(\alpha)}_i$.
Following this way,  Eq.~\eqref{eq5} are reformulated as
\begin{equation} \label{eq6}
\frac{d}{dt}\bm{\eta}=[\mathcal{D}\bm{F}(\bm{s})+\sigma^{(\alpha)} \mathcal{D}
\bm{H}^{(\alpha)}(\bm{s})+\sigma^{(\beta)}\mathcal{D}\bm{H}^{(\beta)}(\bm{s})]\bm{\eta},
\end{equation}
with $\sigma^{(\alpha)}=\varepsilon^{(\alpha)}\lambda^{(\alpha)}_i$,
$\sigma^{(\beta)}=\varepsilon^{(\beta)}\sum_{j=2}^{N}{\mathcal{L}^{(\beta,\alpha)}_{i,j}}$,
$\alpha,\beta=1,2$ and $\alpha\neq\beta$, where $i=2,3,...,N$.
To be stressed, $\sigma^{(\alpha)}$ and $\sigma^{(\beta)}$ encode the
information on the structure of two layers of a duplex network.
We define a modified MSF as the LLE determined from Eq.~\eqref{eq6}, which is
the function of $\sigma^{(1)}$ and $\sigma^{(2)}$ at a given node dynamics and CFs.
In Fig.~\ref{fig3}, $\Lambda(\sigma^{(1)},\sigma^{(2)})=0$ is plotted on the plane
of $\sigma^{(1)}$ and $\sigma^{(2)}$ with the same CFs and node dyanmics as those
in Fig.~\ref{fig2}. Using Fig.~\ref{fig3}, we may determine the stability of CS for
coupled Lorenz oscillators on the duplex network used in Fig.~\ref{fig2} at any
given $\varepsilon^{(1)}$ and $\varepsilon^{(2)}$ as following. Firstly, we calculate
the eigenvalues and eigenvectors, $\lambda_i^{(\alpha)}$ and $\phi_i^{(\alpha)}$
($\alpha=1,2$ and $i=2,3,...,N$) for both layers of the duplex network. Then we calculate
$\sum_{j=2}^{N}{\mathcal{L}^{(2,1)}_{i,j}}$ and
$\sum_{j=2}^{N}{\mathcal{L}^{(1,2)}_{i,j}}$. Secondly, we search for two sets
of $(\varepsilon^{(1)},\varepsilon^{(2)})$ which leads $(\sigma^{(1)},\sigma^{(2)})=
(\varepsilon^{(1)}\lambda^{(1)}_i,\varepsilon^{(2)}\sum_{j=2}^{N}{\mathcal{L}^
{(2,1)}_{i,j}})$ and $(\sigma^{(1)},\sigma^{(2)})=(\varepsilon^{(1)}
\sum_{j=2}^{N}{\mathcal{L}^{(1,2)}_{i,j}},\varepsilon^{(2)}\lambda^{(2)}_i)$,
for $i=2,3,...,N$, to satisfy $\Lambda(\sigma^{(1)},\sigma^{(2)})<0$, respectively.
In the intersection of these two sets of $(\varepsilon^{(1)},\varepsilon^{(2)})$,
CS on the duplex network is stable. Using the modified MSF method on duplex networks,
the stability diagrams of CS on the plane of $\varepsilon^{(1)}$ and $\varepsilon^{(2)}$
in Fig.~\ref{fig2} are reproduced in Fig.~\ref{fig4}. It seems that the stable CS
regions predicted by the modified MSF method are similar to those based on the
synchronization error.

\subsection{Approximation method II}

\begin{figure}
\begin{center}
\includegraphics[width=3.4in]{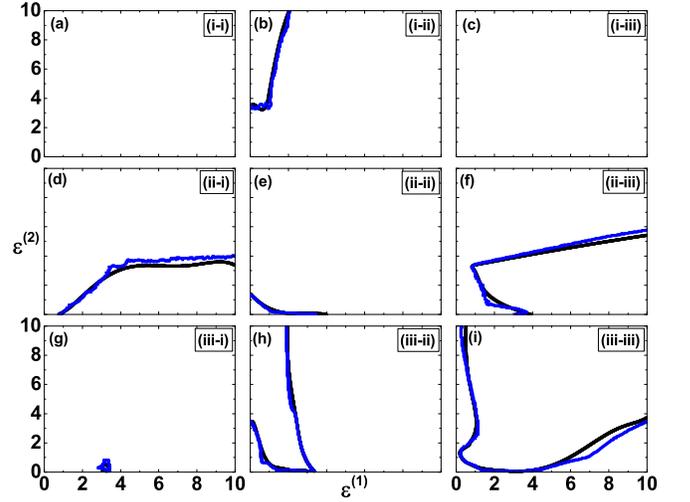}
\end{center}
\caption{\label{fig5}
The stable diagram of CS  predicted by the approximation method II (in blue)
and obtained by the synchronization error (in black).
The CFs are identical to those in Fig.~\ref{fig2}.}
\end{figure}

The MSF method on monolayer networks has shown that the most dangerous transverse
eigenmodes to the stability of CS are those with $\sigma$ close to $\sigma_1$ or $\sigma_2$.
For example, the eigenmode $\phi_2$ with eigenvalue $\lambda_2$ for a type-ii
CF or the eigenmodes $\phi_2$ and $\phi_N$ with eigenvalues $\lambda_2$ and $\lambda_N$
for a type-iii CF. Bearing these facts in mind, we assume that the stability of CS
on duplex networks are mainly determined by four eigenmodes, $\phi^{(1)}_{2}$,
$\phi^{(1)}_{N}$, $\phi^{(2)}_2$, and $\phi^{(2)}_N$. Thereby, the stability of CS
on duplex networks are determined by
\begin{equation} \label{eq7}
\left\{\begin{aligned}
\frac{d}{dt}\bm{\eta}^{(\alpha)}_2&=[\mathcal{D}\bm{F}(\bm{s})+\varepsilon^{(\alpha)}
\lambda^{(\alpha)}_2\mathcal{D}\bm{H}^{(\alpha)}(\bm{s})]\bm{\eta}^{(\alpha)}_2 \\&+
{\mathcal{D}\bm{H}^{(\beta)}(\bm{s})[\mathcal{L}^{(\beta,\alpha)}_{2,2}\bm{\eta}^{(\alpha)}_2
+\mathcal{L}^{(\beta,\alpha)}_{2,N}\bm{\eta}^{(\alpha)}_N]},\\
\frac{d}{dt}\bm{\eta}^{(\alpha)}_N&=[\mathcal{D}\bm{F}(\bm{s})+\varepsilon^{(\alpha)}
\lambda^{(\alpha)}_N\mathcal{D}\bm{H}^{(\alpha)}(\bm{s})]\bm{\eta}^{(\alpha)}_N \\&+
{\mathcal{D}\bm{H}^{(\beta)}(\bm{s})[\mathcal{L}^{(\beta,\alpha)}_{N,2}\bm{\eta}^{(\alpha)}_2
+\mathcal{L}^{(\beta,\alpha)}_{N,N}\bm{\eta}^{(\alpha)}_N]},
\end{aligned}\right.
\end{equation}
with $\alpha,\beta=1,2$ and $\alpha\neq\beta$.
When the LLE determined by Eq.~\eqref{eq7} are negative, the CS on duplex
networks is stable. Through the truncation approximation, Eq.~\eqref{eq5},
the $N-1$ $m$-dimensional ODEs, are reduced to $4m$-dimensional ODEs.
For large $N$, the approximation here greatly simplifies the problem of stability
analysis on CS on duplex networks. Fig.~\ref{fig5} shows the region on the plane
of $\varepsilon^{(1)}$ and $\varepsilon^{(2)}$ predicted by Eq.~\eqref{eq7}.
Clearly, the boundary of the stable CS region is almost the same as those obtained
by the synchronization error.

\subsection{Approximation methods applied on systems with larger sizes}

\begin{figure}
\begin{center}
\includegraphics[width=3.4in]{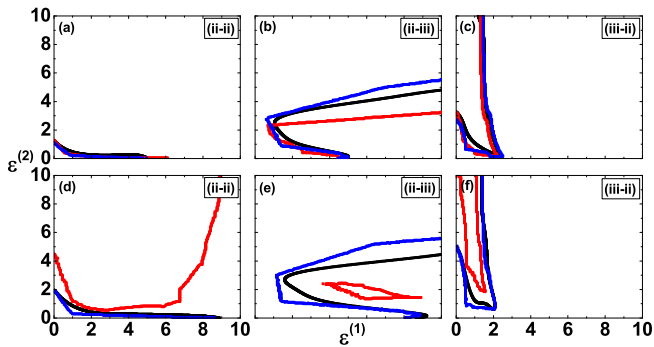}
\end{center}
\caption{\label{fig6}
The stable diagram of CS predicted by the approximation method I (in red),
the approximation II (in blue), and obtained by the synchronization error (in black).
The top panel for $N=20$ and the bottom panel for $N=100$.
Each layer for duplex networks is a regular random network with $k=4$.
For different combinations of CFs,
(a) and (d) $\mathcal{D}\bm{H}^{(1)}_{1,1}=1$ and $\mathcal{D}\bm{H}^{(2)}_{2,2}=1$.
(b) and (e) $\mathcal{D}\bm{H}^{(1)}_{1,1}=1$ and $\mathcal{D}\bm{H}^{(2)}_{1,2}=1$.
(c) and (f) $\mathcal{D}\bm{H}^{(1)}_{1,2}=1$ and $\mathcal{D}\bm{H}^{(2)}_{2,1}=1$.}
\end{figure}

As shown in Figs.~\ref{fig4} and \ref{fig5}, the two approximation methods seem
to work quite well on predicting the stability of CS on duplex networks with $N=6$.
Then we wonder the validity of the approximation methods for duplex networks with
large $N$. We consider $N=20$ and $N=100$, respectively. In both examples, each
layer of duplex networks is modelled by a regular random network with degree
$k=4$. The top panel and the bottom panel in Fig.~\ref{fig6} show stable CS
regions calculated by the synchronization error (in black), the approximation
method I (in red), and the approximation method II (in blue) for $N=20$ and
$N=100$, respectively.
Clearly, the approximation method II works much better than the approximation
method I. Both approximation methods work pretty well on predicting the stable
CS region at $N=20$. At $N=100$, the stable CS region predicted by the approximation
method I deviates largely from those acquired by the synchronization error while the
stable CS region predicted by the approximation method II only shows a little mismatch
with those obtained by the synchronization error. Fig.~\ref{fig6} also shows that the
approximation method I (or the approximation method II) predicts a smaller (or larger)
stable region than those obtained by the synchronization error. In addition, we
have checked the validity of two approximation methods on the stability of
CS on duplex networks in which each layer is modelled by an Erd\"os-R\'enyi network or
a scale free network and found the results similar to Fig.~\ref{fig6}.

\section{Discussion and conclusions}

To conclude, we studied the stability of the complete synchronization
in coupled systems upon duplex networks in this work.
We found that the combinations of CFs on different layers have strong effects
on the stability of CS. We propose two approximation methods to predict the
stable CS region on the $\varepsilon^{(1)}-\varepsilon^{(2)}$ plane.
In the approximation I, we assumed that all transverse eigenmodes of each layer
have the same contribution to the stability of CS and, as a result, we could
introduce a modified MSF. In the approximation II,
we only considered the contributions of a few most unstable eigenmodes in each layer
to the stability of CS. We found that both approximation methods could work pretty
well for small networks. For networks with large size, the approximation II could
work much better than the approximation I.
We expect our work should be instructive for the future works on
synchronization on multiplex networks, such as optimal synchronization,
which attracted much attention on monolayer networks \cite{don05,zhou06,ska14},
as well as the system of opinion dynamics, information spreading and so on.

\section{Acknowledgements}
This work is supported by the National Natural Science Foundation of China
under grant Nos. 11575036 and 11505016.

\end{document}